\RequirePackage{fix-cm}
\documentclass[smallextended]{svjour3}       
\smartqed  
\usepackage{graphicx}
\begin{document}

\title{First direct measurement of positronium hyperfine splitting with sub-THz light
}


\author{T. Yamazaki         \and
        A. Miyazaki \and T. Suehara \and T. Namba \and S. Asai \and T. Kobayashi \and H. Saito \and Y. Urushizaki \and I. Ogawa \and T. Idehara \and S. Sabchevski 
}


\institute{T. Yamazaki \at
              Department of Physics, Graduate School of Science, and International Center for Elementary Particle Physics, university of Tokyo, Hongo, Bunkyo-ku, Tokyo 113-0033, Japan \\
              Tel.: +81-3-3815-8384\\
              Fax: +81-3-3814-8806\\
              \email{yamazaki@icepp.s.u-tokyo.ac.jp}           
           \and
           A. Miyazaki \and T. Suehara \and T. Namba \and S. Asai \and T. Kobayashi \at
              Department of Physics, Graduate School of Science, and International Center for Elementary Particle Physics, University of Tokyo \\
	   \and
           H. Saito \at
	      Graduate School of Arts and Sciences, University of Tokyo \\
           \and
           Y. Urushizaki \and I. Ogawa \and T. Idehara \at
              Reserch Center for Development of Far-Infrared Region, University of Fukui \\
           \and
           S. Sabchevski \at
              Bulugarian Academy of Sciences
}

\maketitle

\begin{abstract}
Positronium is an ideal system for the research of the bound state QED. The hyperfine splitting of positronium (Ps-HFS, about 203 GHz) is an important observable but all previous measurements of Ps-HFS had been measured indirectly using Zeeman splitting. There might be the unknown systematic errors on the uniformity of magnetic field. We are trying to measure Ps-HFS directly using sub-THz radiation. We developed an optical system to accumulate high power (about 10 kW) radiation in a Fabry-P$\acute{\rm{e}}$rot resonant cavity and observed the positronium hyperfine transition for the first time.
\keywords{Positronium \and Hyperfine splitting}
\end{abstract}

\section{Introduction}
\label{intro}
Positronium (Ps), the bound state of an electron and a positron, is an ideal system for the research of the bound state quantum electrodynamics (QED). The spin triplet (1$^3S_1$) state of Ps, ortho-positronium (o-Ps), has long lifetime (142 ns) and decays into three $\gamma$ rays. On the other hand, the spin singlet (1$^1S_0$) state of Ps, para-positronium (p-Ps), has very short lifetime (125 ps) and decays into two $\gamma$ rays promptly. The difference of the energy level between o-Ps and p-Ps is called hyperfine splitting of positronium (Ps-HFS) and is significantly larger (about 203 GHz) than that of hydorogen atom (about 1.4 GHz). The Ps-HFS is a good measure to validate the bound state QED and sensitive to the new physics beyond the Standard Model.

The precise measurements have been performed in 1970's and 1980's and the measured value is 203.388 65(67) GHz\cite{Mills}\cite{Ritter}. All previous experiments employed static magnetic field (about 1 T) and Ps-HFS has been measured indirectly using Zeeman splitting (about 3 GHz). We prepare a new experiment to measure the Ps-HFS directly using sub-THz radiation.

Direct hyperfine transition has never observed so far because the probability of spontaneous emission (Einstein A coefficient) of Ps-HFS is $3\times10^{-9} {\rm sec}^{-1}$, which is smaller than that of o-Ps decay by 14 orders of magnitude. Therefore, high power (over 10 kW) sub-THz radiation is necessary to drive enough stimulated transitions from o-Ps to p-Ps. Furthermore, the radiation source should be frequency-tunable by about 10 GHz to obtain the resonance curve. We propose to use a gyrotron as radiation source and accumulate its output in a Fabry-P$\acute{\rm e}$rot resonant cavity.

\section{Experimental setup}
\label{sec:setup}
\begin{figure}
  \includegraphics[width=0.75\textwidth]{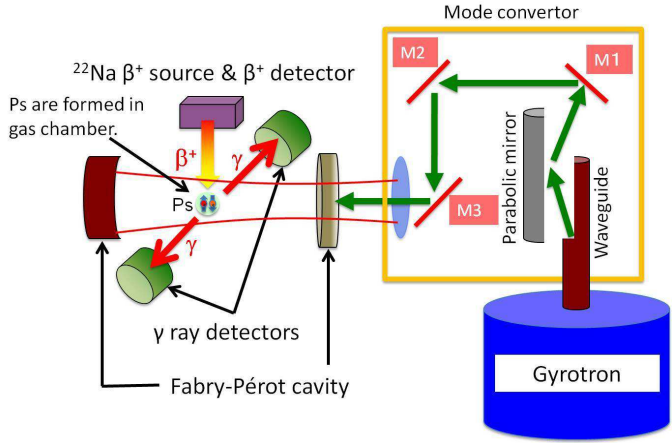}
\caption{Experimental setup}
\label{fig:setup}       
\end{figure}

Figure \ref{fig:setup} shows an schematic of our experimental setup. It consists of an optical system to accumulate high power sub-THz radiation, Ps assembly, and $\gamma$-ray detectors.

\subsection{Optical system}
\label{sec:optics}
The optical system consists of a gyrotron (radiation source), a mode converter, and a Fabry-P$\acute{\rm e}$rot resonant cavity.

The gyrotron is an electron cyclotron maser and a novel high power radiation source for sub-THz to THz frequency region. The electrons are emitted from the electron gun, concentrated and rotated as cyclotron motion in the magnetic field. The electrons enter the cavity of the gyrotron and then the energy of the cyclotron motion is converted to that of the electromagnetic wave in the cavity if we tune the strength of the magnetic field so that the cyclotron frequency is slightly smaller than the cavity resonant frequency.

\begin{figure}
  \includegraphics[width=0.50\textwidth]{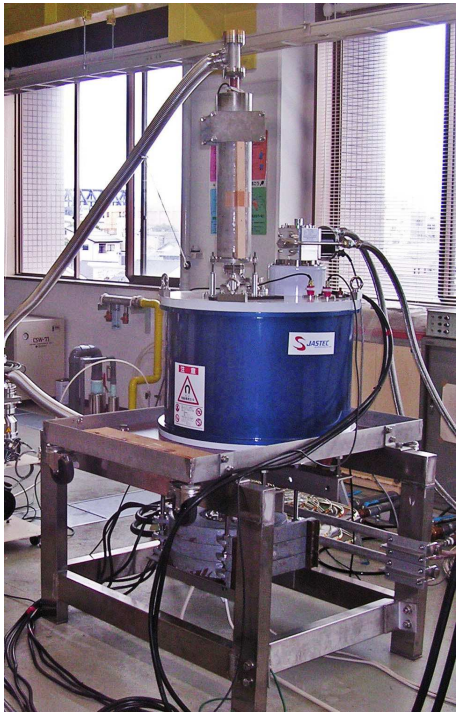}
\caption{Gyrotron FU CW V}
\label{fig:gyrotron}       
\end{figure}

A gyrotron Gyrotron FU CW V (Fig. \ref{fig:gyrotron}) are used, in which the wave mode of its output radiation is TE$_{03}$ mode. The radiation frequency is 202.9 GHz and the stable power of about 300 W is obtained.

The power of the gyrotron output radiation is high (about 300 W), but it is far from our requirements (over 10 kW). Therefore, we have to accumulate the radiation in the Fabry-P$\acute{\rm e}$rot resonant cavity. It is necessary to convert the wave mode into TEM$_{00}$ "Gaussian beam" to match with the wave mode in the cavity.

\begin{figure}
  \includegraphics[width=0.75\textwidth]{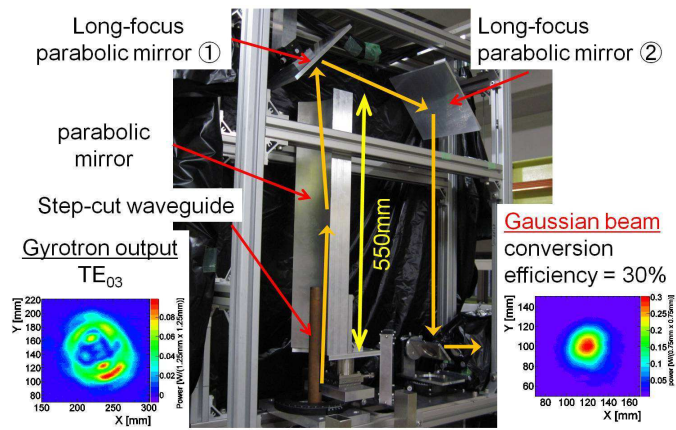}
\caption{Mode converter. Power profiles of beam are also shown. TE$_{03}$ is just after the output of gyrotron and Gaussian beam is after the mode converter.}
\label{fig:converter}       
\end{figure}

Figure \ref{fig:converter} shows the mode converter which consists of a step-cut waveguide and three large parabolic mirrors. The step-cut waveguide and the first parabolic mirror convert the gyrotron output (TE$_{03}$) into polarized plain wave (bi-Gaussian beam). The following two parabolic mirrors are arranged to make shape of beam and convert it to Gaussian beam. The conversion efficiency is about 30\% because the wave mode of the gyrotron output is not perfect TE$_{03}$ waveguide mode.

\begin{figure}
  \includegraphics[width=0.75\textwidth]{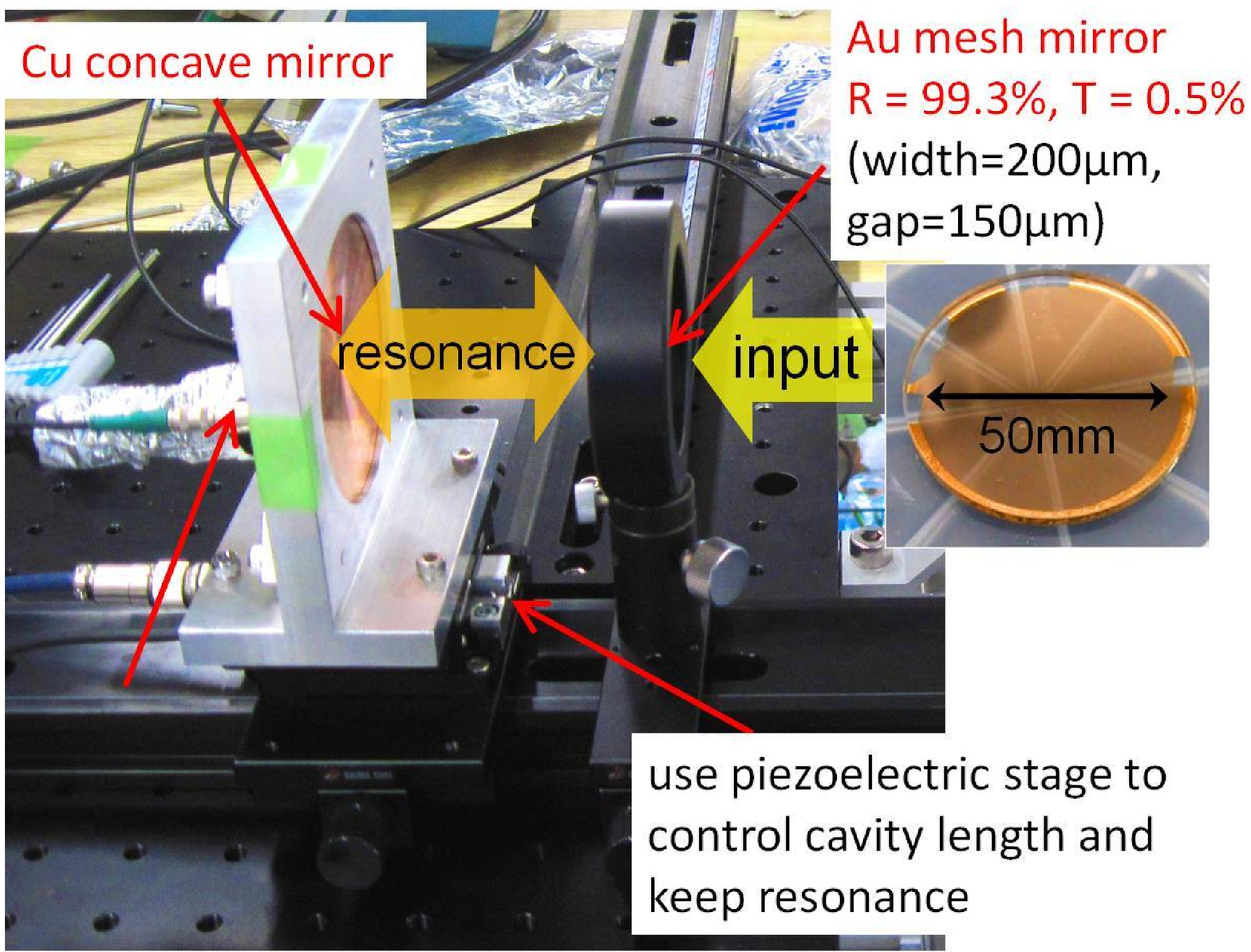}
\caption{Fabry-P$\acute{\rm e}$rot resonant cavity}
\label{fig:fabry-perot}       
\end{figure}

The converted Gaussian beam which enters the Fabry-P$\acute{\rm e}$rot resonant cavity makes many round-trips in the cavity. The Fabry-P$\acute{\rm e}$rot resonant cavity consists of two high reflective mirrors (Fig. \ref{fig:fabry-perot}). We use a gold-mesh mirror on the input side of the cavity and a copper concave mirror on the other side. The reflectivity of the gold-mesh mirror is very high (99.3\%) and the transmittance is reasonable (0.3\%). It is also important that the gold-mesh mirror never disturbs Gaussian beam. Piezo stage is used to control the interval of two mirrors.

\begin{figure}
  \includegraphics[width=\textwidth]{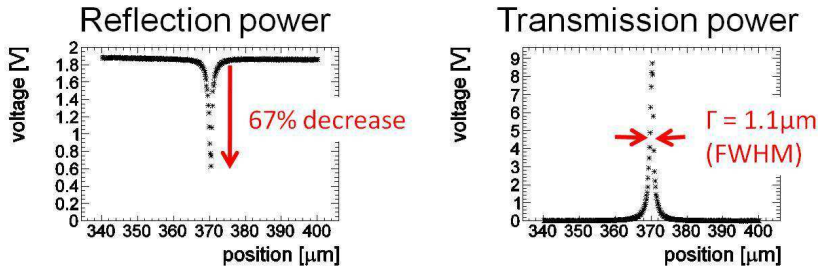}
\caption{(a) Reflection power of the cavity (b) Transmission power of the cavity}
\label{fig:FP-spec}       
\end{figure}

The two important characteristics of the cavity are ``coupling $C$'' and ``finesse $\mathcal{F}$''. The input coupling can be estimated from the decrease of the reflection power of the cavity when resonance. The power reflected by the cavity input are shown in Fig. \ref{fig:FP-spec} (a), and the coupling 67\% is obtained. Finesse can be written as $\mathcal{F} = 2\pi N_{\rm round} = \lambda/2\Gamma$, where $N_{\rm round}$ is the number of round-trips in the cavity, $\lambda$ is the wavelegth of the sub-THz radiation (about 1.5 mm), and $\Gamma$ is the sharpness of the resonance (FWHM). Power transmittance through the cavity is shown in Fig. \ref{fig:FP-spec} (b) and clear resonance ($\Gamma = 1.1 \mu{\rm m}$) is observed. $\mathcal{F}$ is therefore about 650 and $N_{\rm round}$ is about 100. As a result, the gain of the cavity is $G = C\times2N_{\rm round} = 130$.

The accumulated power of the optical system is
\begin{equation}
P_{\rm acc} = P_{\rm gyro} \times \epsilon_{\rm conv} \times G = 300 [{\rm W}] \times 0.3 \times 130 = 12 [{\rm kW}],
\end{equation}
which higher than our requirement (over 10 kW).

\subsection{Ps assembly and $\gamma$-ray detectors}
\begin{figure}
  \includegraphics[width=\textwidth]{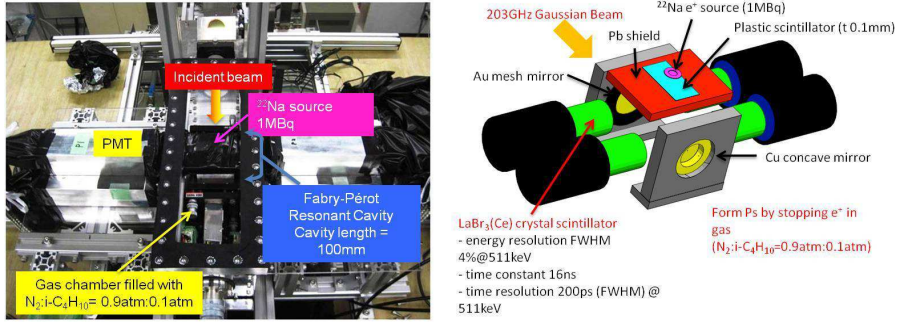}
\caption{Ps assembly and $\gamma$-ray detectors}
\label{fig:assembly}       
\end{figure}

Figure \ref{fig:assembly} shows a schematic and a picture of the Ps assembly filled with the gas and $\gamma$-ray detectros.
A positron emitted from $^{22}$Na positron source (1 MBq) stops in the gas and forms the positronium. The gas is mixture of nitrogen of 0.9 atm and i-C$_4$H$_{10}$ of 0.1 atm. i-C$_4$H$_{10}$ is used to quench the slow positron. High power radiation in the Fabry-P$\acute{\rm e}$rot resonant cavity drives stimulated transition from o-Ps to p-Ps and the p-Ps decays into two $\gamma$ rays promptly. The $\gamma$ rays are detected by surrounding $\gamma$-ray detectors as shown in Fig. \ref{fig:assembly}. Therefore, the transition signal has a distinctive feature of ``two back-to-back 511 keV $\gamma$ rays with long lifetime (about 142 ns)''. LaBr$_3$ crystal scintillators are used as the $\gamma$-ray detectors because their good energy resolution (FWHM = 4\% at 511 keV) is of advantage to find 511 keV peak in the energy spectrum and their fast response (time constant = 16 ns) is suited to high statistics experiments. In order to obtain the information of the lifetime of a positronium, the timing of the positron emission is tagged with a plastic scintillator. The time difference between the positron emission and $\gamma$-ray detection corresponds to the lifetime of positronium. 

Run time on the resonance is $5\times10^{5}$ sec. The trigger of data aquisition is as follows: At least two signals from the LaBr$_3$ scintillators are coincident within 20 ns, and then when this coincidence is within $-$100 ns to 1100 ns of the timing of the plastic scintillators. The trigger rate was about 480 Hz. During the data aquisition, the average accumulated power in the Fabry-P$\acute{e}$rot resonant cavity was 11.6$^{+3.0}_{-2.6}$ kW, which was measured by a pyroelectric detector.

\section{Data analysis}
\begin{figure}
  \includegraphics[width=0.75\textwidth]{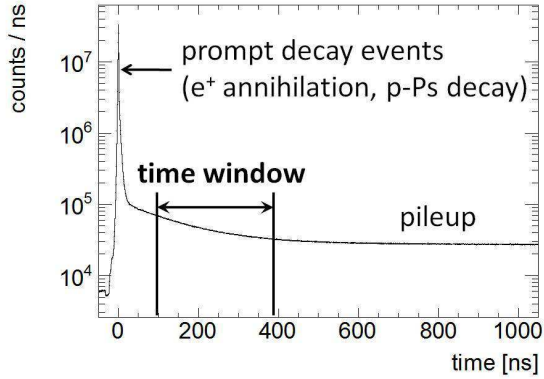}
\caption{The time difference between the plastic scintillator and the coincidence signals of the LaBr$_3$ scintillators.}
\label{fig:tspec}       
\end{figure}

Figure \ref{fig:tspec} shows the time difference between the plastic scintillator and the coincidence signals of the LaBr$_3$ scintillators. A sharp peak from prompt annihilation is followed by the exponential curves of o-Ps and transition signal and then the constant accidental spectrum. Fitted lifetime is 136 ns. A time window from 100 ns to 400 ns is required to reject prompt events and pileup events and improve S/N significantly.

\begin{figure}
  \includegraphics[width=\textwidth]{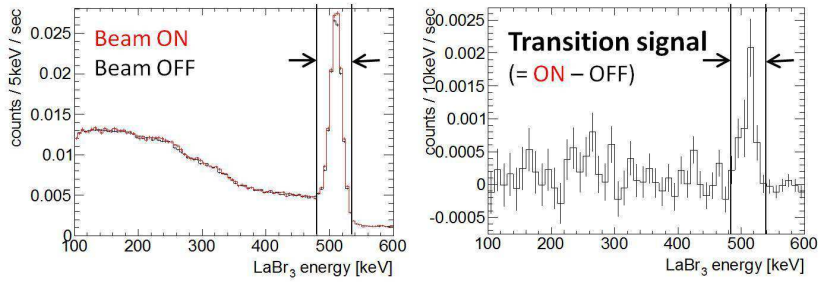}
\caption{The energy spectra after all event selections.}
\label{fig:transition}       
\end{figure}

Figure \ref{fig:transition} shows the energy spectra after all event selections are applied. Spectrum with "OFF" power is subtracted from that with "ON" power to draw out the transition. Figure \ref{fig:transition} shows the subtracted spectrum and transition signal is clearly observed. Gyrotron power has the pulse structure. Power is "ON" during 60 msec for every 200 msec (5Hz, duty factor 30\%). Pulse is fast riging enough.

The systematic errors are summarized in Table \ref{tab:systematic}. The quadrature sum is 8.8\% of the transition signal.
This is the first observation of direct hyperfine transition of positronium. Einstein A coefficient is estimated from the amount of the transition signal and the average accumulated power during the data acquisition. A is obtained $3.9^{+1.8}_{-1.7} \times 10^{-8} [{\rm s}^{-1}]$, which is consitent with its theoretical value ($3.37 \times 10^{-8} [{\rm s}^{-1}]$).

\begin{table}
\caption{Summary of the systematic errors}
\label{tab:systematic}       
\begin{tabular}{lc}
\hline\noalign{\smallskip}
\multicolumn{1}{c}{Source} & Systematic error  \\
\noalign{\smallskip}\hline\noalign{\smallskip}
LaBr$_3$ energy resolution & 3.0\% \\
Origin of the time spectrum & 0.3\% \\
positron tag rate & 7.7\% \\
normalization & 3.1\% \\
\noalign{\smallskip}\hline
\end{tabular}
\end{table}

\section{Summary and plan}
We prepare an experiment to measure the hyperfine splitting of the positronium directly. High power sub-THz radiation is necessary to drive enough stimulated transition between o-Ps to p-Ps because the probability of the hyperfine transition is extreamly small compared with that of the o-Ps decay. We have already developed a new optical system to accumulate sub-THz radiation and achieved accumulated power of 10 kW. The direct transition between o-Ps to p-Ps was discovered for the first time with the optical system. The amount of the observed transition was consistent with its theoretical calculation.

In order to measure the Ps-HFS, we are now developing a frequency-tunable gyrotron and upgrading the Ps assembly and the $\gamma$-ray detection system to improve statistics and S/N. We plan to make the first measurement of the Ps-HFS in a year.




\end{document}